\documentclass[aps,prb,twocolumn,10pt]{revtex4-1}
\usepackage{amssymb,amsmath}
\usepackage{graphicx}
\usepackage{dcolumn}
\usepackage{hyperref}
\usepackage{color}
\begin{document}

\title{Water Bottle Flipping Physics}
\author{P.J. Dekker$^1$, L.A.G. Eek$^1$, M.M. Flapper$^1$, H.J.C. Horstink$^1$, A.R. Meulenkamp$^1$, J. van der Meulen$^1$, E.S. Kooij$^{1,2}$, J.H. Snoeijer$^{1,3}$, A. Marin$^{1,3}$}
\affiliation{
$^1$~Faculty of Science and Technology, University of Twente, P.O. Box 217, 7500 AE
Enschede, The Netherlands\\
$^2$~Physics of Interfaces and Nanomaterials, MESA+ Institute for Nanotechnology,
University of Twente, P. O. Box 217, 7500 AE Enschede, The Netherlands\\
$^3$~Physics of Fluids Group, J.M. Burgers Center
for Fluid Dynamics, University of Twente, P.O. Box 217, 7500 AE
Enschede, The Netherlands
}

\date{\today}

\begin{abstract}
The water bottle flipping challenge consists of spinning a bottle, partially filled with water, and making it land upright. It is quite a striking phenomenon, since at first sight it appears rather improbable that a tall rotating bottle could make such a stable landing. Here we analyze the physics behind the water bottle flip, based on experiments and an analytical model that can be used in the classroom. Our measurements show that the angular velocity of the bottle decreases dramatically, enabling a nearly vertical descent and a successful landing. The reduced rotation is due to an increase of the moment of inertia, caused by the in-flight redistribution of the water mass along the bottle. Experimental and analytical results are compared quantitatively, and we demonstrate how to optimize the chances for a successful landing. 
\end{abstract}

\maketitle

\section{Introduction}

In May 2016, a senior high school student called Michael Senatore enters a stage carrying a partially filled bottle of water. He is participating in the school's annual talent show and the auditorium is packed. There is music playing in the background as he approaches the center of the stage in a funny way (swagger move). Suddenly he gets serious, stands straight, focuses on a table standing in front of him and throws the bottle in the air with a spin. The bottle flipped once and landed standing perfectly upright on a table. This brings down the house and the students bursts in wild cheers. Everything was filmed with a smartphone camera \cite{USAtoday}.
Within weeks, this 30 seconds clip becomes viral on the internet and kids around the globe are seen attempting the ``water bottle flipping challenge'', as is it known nowadays \cite{NYtimes,Vox}. Michael Senatore ended up selling the famous bottle for 15.000\$ (or at least one signed by him) \cite{reuters}. 

Rotational physics often involves rather counterintuitive phenomena like the rotation of cats in free-fall~\cite{spincats} or Olympic divers~\cite{momentumviolators}, but the remarkable water bottle flip is no exception. Yet, the flip offers an original and very insightful illustration of the fundamental principles of rotational mechanics. In Fig.~\ref{fig:images}(a) (and in the supplementary video footage\cite{supplVideo}) we present a series of snapshots of a successful flip. At first sight it appears rather improbable that a tall rotating bottle could land stably in upright position. \textcolor{black}{After all, once released, the bottle’s angular momentum with respect to the center of mass must be conserved.} \textcolor{black}{For a rigid body rotation around a principle axis, the conservation of angular momentum implies a rotation with a constant angular velocity, making a smooth landing rather unlikely.} However, the sloshing of the water leads to a redistribution of the mass along the bottle. This change of mass distribution is clearly visible in the top row of Fig.~\ref{fig:images} (while the bottle is in the air), and will increase the moment of inertia. Conservation of angular momentum then implies a \emph{decrease} of the rotational velocity -- leaving the impression of the bottle being suspended horizontally in the air for a moment. When performed successfully, the flip ends with a nearly vertical descent that is followed by a smooth landing. 

%\textcolor{red}{Rotational physics allow for several non-intuitive examples that can be used in the classroom. Despite of the conservation of angular momentum during free falls, cats manage to rotate along their body axis in air to fall on their feet without generating a net spin\cite{spincats} and Olympic divers operate their body to make a precise numbers of turns before carefully and timely adjusting their final position \cite{momentumviolators}.}

%
\begin{figure*}[t]
% \begin{center}
% \includegraphics[width=\textwidth]{Fig1.png}
\includegraphics[width=.95\textwidth]{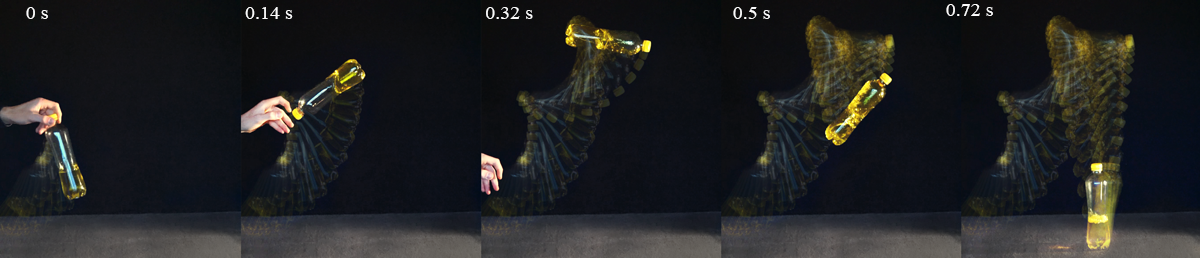}
\includegraphics[width=.95\textwidth]{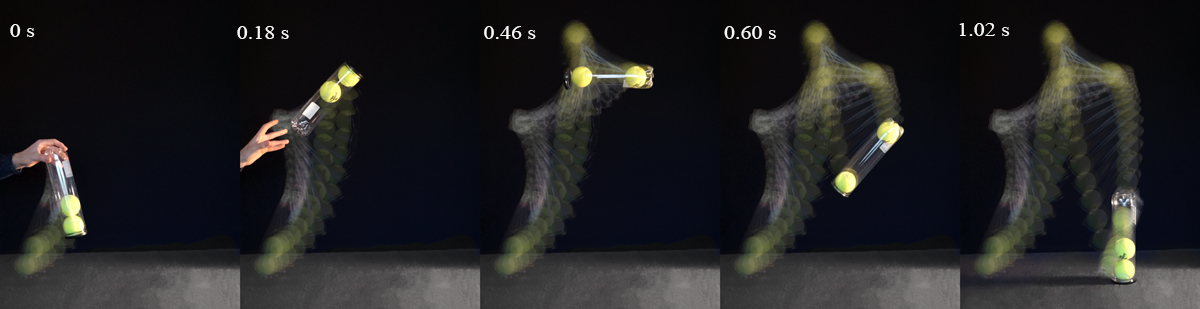}
\caption{Compositional photographs of a water bottle flip (top), and a tennis bottle flip (bottom). In both cases the redistribution of mass inside the bottle leads to an increase of the moment of inertia -- slowing down its rotational speed and allowing for a near-vertical descent.} \label{fig:images}
% \end{center}
\end{figure*}
% %
% \begin{figure}[ht!]
% % \begin{center}
% \includegraphics[width=.5\textwidth]{Fig1.png}
% \caption{Compositional photographs of a water bottle flip (a), and a tennis bottle flip (b). In both cases the redistribution of mass inside the bottle leads to an increase of the moment of inertia -- slowing down its rotational speed and allowing for a near-vertical descent.} \label{fig:images}
% % \end{center}
% \end{figure}
% %

In this paper we demonstrate how the water bottle flip can be used in the classroom. In Sec.~\ref{sec:exp} we show how the complex dynamics of the bottle can be imaged in experiments, and how it can be analyzed by separating the motion in the translation of the center of mass and a rotation around the center of mass. Since the physics of water sloshing is highly complex in itself, we present an alternative that is more suitable for analysis: The ``tennis bottle flip", shown in Fig.~\ref{fig:images}(b) (and in the supplementary video footage\cite{supplVideo}). In this system the water is replaced by two tennis balls -- indeed, the successful tennis bottle flip clearly demonstrates that the redistribution of mass is the physical ingredient behind the flip. Subsequently, in Sec.~\ref{sec:model} we show how the flip can be described by a theoretical model, even allowing quantitative comparison to experiments. 

Based on the observations and modeling, we close by addressing an important question that arises when attempting a water bottle flip challenge (Sec.~\ref{sec:optimal}): Why is there an optimal amount of water in the bottle? Millions of flippers seem to disagree on the precise value, but they do agree that the  optimal filling fraction should be between 1/4 and 1/3 of the total height of the bottle. Can we explain these values from mechanical principles?

\begin{figure}
\begin{center}
\includegraphics[width=.5\textwidth]{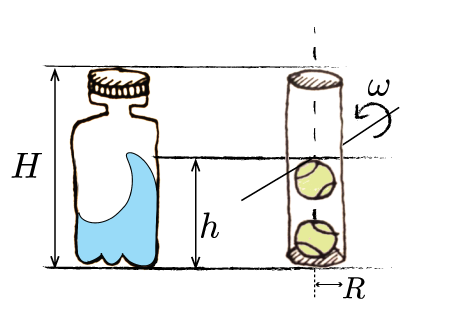}
\caption{Sketch of the geometrical parameters for the water bottle and tennis bottle. The total height of the bottle is $H$, while the distribution of water/balls is indicated as $h$. The axis of rotation is indicated as well.} 
\label{fig:sketch}
\end{center}
\end{figure}
\begin{figure*}[t]
% \begin{center}

% \includegraphics[height=0.25\textwidth]{figFixed.pdf}
% \includegraphics[height=0.25\textwidth]{figWaterExperiment.pdf}
% \includegraphics[height=0.25\textwidth]{figTennisExperiment.pdf}
 \includegraphics[width=0.9\textwidth]{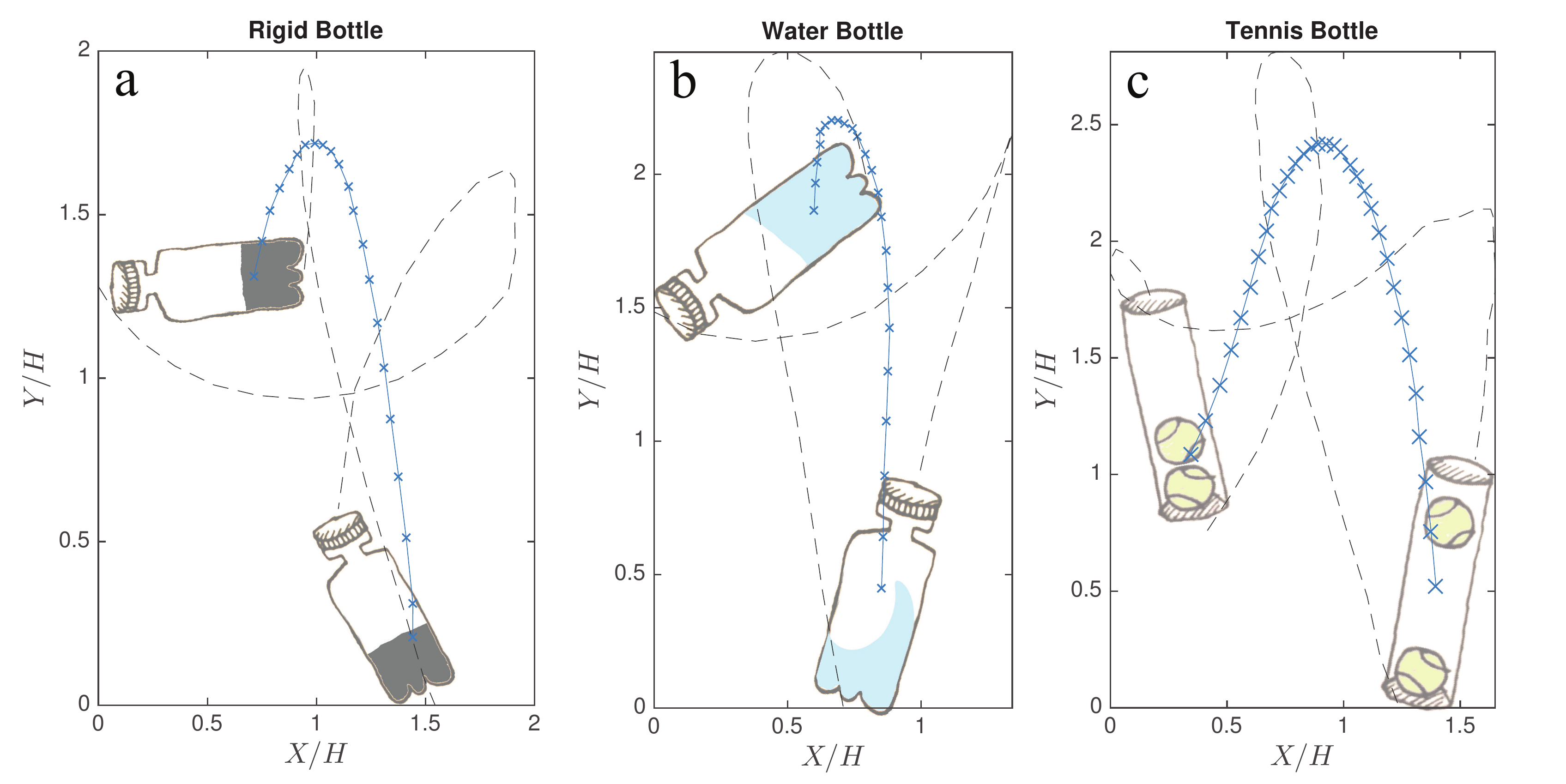}
\caption{Analysis of the motion during the flip for (a) a bottle with immobilized mass, (b) a water bottle, (c) a tennis bottle. In all panels the gray dashed lines represent the complex trajectories of the top and bottom of the bottle. The continuous blue lines describe the center of mass motion, which to a good approximation is found to follow a parabolic trajectory. The values of $H$ used in the shown experiments are $H$ = 23 cm for the rigid and water bottles and 28 cm for the tennis bottle.} 
\label{fig:trajectories}
% \end{center}
\end{figure*}
\section{Experiments}\label{sec:exp}

We start by describing how to visualize the dynamics during a bottle flip, and how to analyze the resulting motion. The experiment is designed along the classical approach for the dynamics of extended bodies\cite{Resnick,Landau}: The motion is decomposed into a \emph{translation} of the center of mass and a \emph{rotation} around the center of mass. This decomposition is natural since the only external force is gravity, which exerts no torque around the center of mass. As such, the water bottle flip serves as a prime example of conservation of angular momentum. 

Apart from the water bottle and the tennis bottle shown in Fig.~\ref{fig:images}, we will also consider a ``rigid bottle" that contains an immobilized mass. The rigid bottle serves two purposes: to verify if we recover the usual rigid body rotation, and to highlight the importance of the movable mass for a successful bottle flip.

%\textcolor{blue}{The aim of the experiments is to track the whole bottle trajectory and its center of mass. The bottle's top and bottom can be easily tracked in the experiments. However, finding the location of the CM for a flipping water bottle is not such a  simple task. An strategy to find the CM indirectly is proposed in the following section \ref{subsec:setup}. Nonetheless, in order to have a flipping system in which we can determine the CM directly from the experiments, we propose two model bottle flippers. The first one is a \emph{rigid body}, i.e. a bottle in which mass cannot redistribute during the flip. We call this system the \emph{rigid bottle}. The second one is a tennis can with capacity for three tennis balls, filled with only two tennis balls. We call this system the \emph{tennis bottle}.}

\subsection{Experimental setup and analysis}\label{subsec:setup}

The experimental setup used in this study consists of a black background, a lamp for illumination, and a digital camera $\alpha$-6000. Each experimental run (or bottle flip) takes roughly 1 second. Here we recorded the films at 50 frames per second, 1/1600 s of shutter time and 2 megapixels resolution. Our recommendation is to use a minimum of 20 frames per second to gather enough data points, a maximum shutter time of 1/200 s to avoid blur in the moving bottle and a minimum resolution of 1 megapixels (most smartphone-cameras satisfy such requirements nowadays). We typically ran 10 successful flips per bottle type with the same fillings, and select the cleanest landings among them for analysis.

The rotational motion is quantified by the angular velocity $\omega = d\theta/dt$. This quantity can be measured by tracing the top and bottom of the bottle on the videos. Another key ingredient of the analysis is to determine the motion of the center of mass of the total system. For the rigid bottle, the center of mass obviously remains at a fixed position along the bottle for all times. However, it is rather difficult to accurately determine the center of mass of the sloshing water -- from the images one cannot infer the precise distribution of water inside the bottle. Here we simply proceed by an approximate analysis that is detailed in Sec.~\ref{subsec:modelwater}, based on the maximum height of the water mass along the bottle. This complexity of the water bottle [Fig.~\ref{fig:images}(a)] is our prime motivation for introducing the tennis bottle [Fig.~\ref{fig:images}(b)]. Namely, the exact positions of the tennis balls are easily determined. Subsequently, the center of mass is obtained by taking the mass-weighted average of the positions of the two balls and of the bottle's center.

In summary, the experimental measurements consist of tracking the top and bottom of the bottle on each frame, and the following additional points to determine the center of mass: (a) Water bottle: the maximum height of the sloshing water $h$ on each frame (see Fig.~\ref{fig:sketch}). (b) Tennis bottle: the position of the tennis balls for each frame during the flip. The acquired digital images were imported into a computer and the tracking was performed manually using ImageJ \cite{ImageJ}, simply using the point tool with auto-measure. The data was then processed using MATLAB \cite{MATLAB}. All manually tracked sets of data were filtered using smoothing splines (with a low smoothing factor of 0.99) to reduce the user-induced bias and noise.

The experiments presented below were performed with a water bottle of mass $m_b=$ 25 g, height $H=$ 23 cm. The filling fraction used was 0.39. For the tennis bottle experiments, a tennis ball has a mass of 58 g, and the tennis bottle has a mass of 48 g and a height of 28 cm and a radius of 3.7 cm.

\subsection{Results}

Figure~\ref{fig:trajectories} shows typical trajectories obtained from our experiments on the rigid bottle (panel a), the water bottle (panel b), and the tennis bottle (panel c). The various curves respectively trace out the edges of the bottle (gray lines) and the center of mass position (blue lines). In all cases, the center of mass follows the expected parabolic trajectory associated to a free-falling motion. The parabola is most convincingly observed for the rigid bottle and tennis bottle [Fig.~\ref{fig:trajectories}(a,c)]. In these experiments the center of mass was indeed accurately determined, while this measurement was more approximate for the sloshing of water [Fig.~\ref{fig:trajectories}(b)]. 

\begin{figure}[b]
\begin{center}
\includegraphics[width=.45\textwidth]{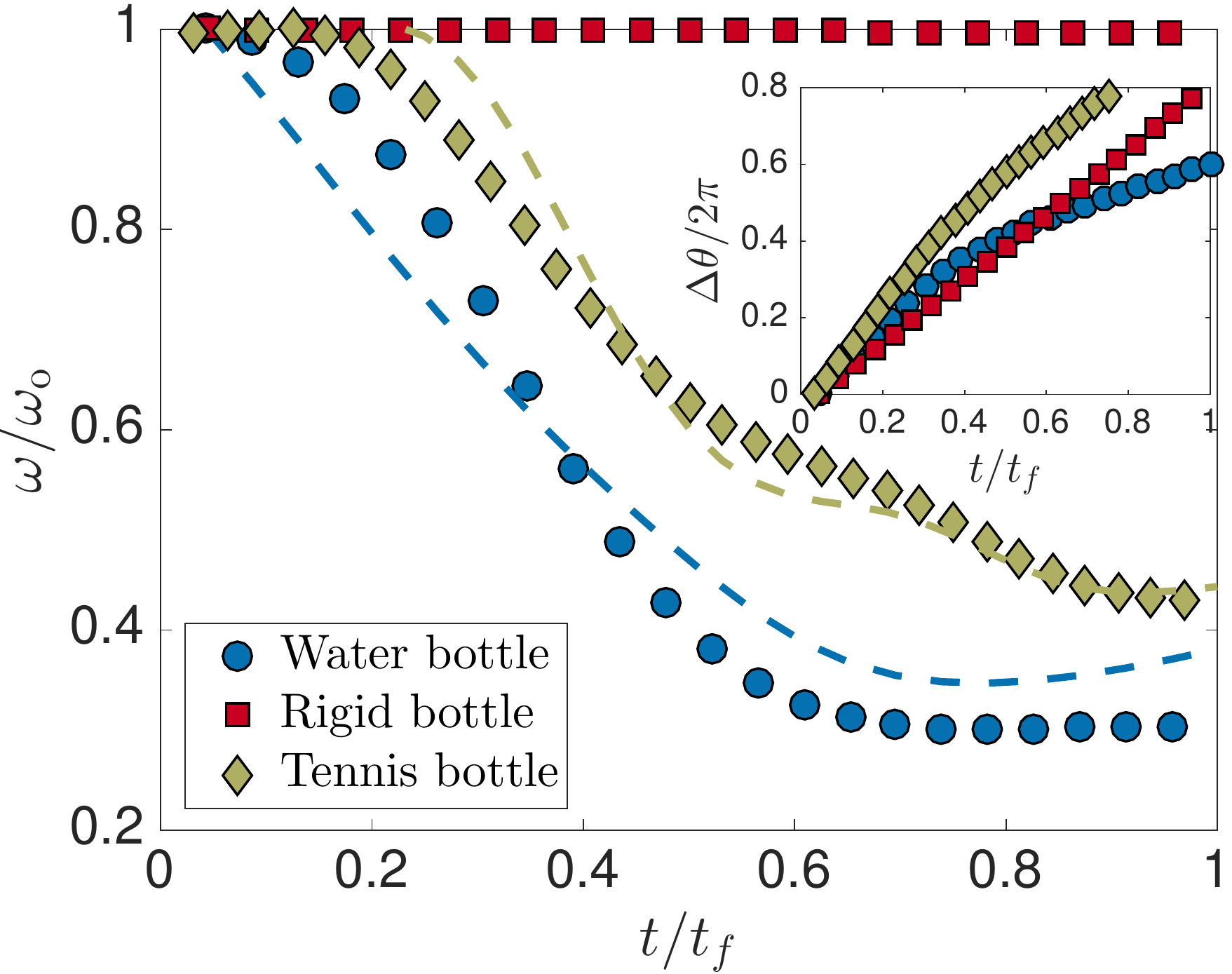}
\caption{Angular velocity $\omega$ as a function of time $t$, respectively normalized by the initial $\omega_0$ and the final time $t_f$. The datasets correspond to the bottle with an immobilized mass (red), the water bottle (blue) and the tennis bottle (yellow). Dashed lines correspond to the model described in Sec.~\ref{sec:model}. Inset: The angle $\Delta\theta=\theta(t)-\theta(0)$ versus $t$, from which the main figure was derived.}
%\textcolor{red}{replace $t_{max}$ by $t_f$ in figure.} } 
\label{fig:angular}
\end{center}
\end{figure}

Our prime interest, however, lies in the rotational aspects of the motion. It is clear from Fig.~\ref{fig:trajectories} that the rotation of the rigid bottle is very different from both the water bottle and the tennis bottle. This is further quantified by considering the angular velocity $\omega = d\theta/dt$, where the angle $\theta(t)$ describes the orientation of the bottle over time. The raw data of $\Delta\theta(t) = \theta(t)-\theta(t=0)$ is shown in the inset of Fig.~\ref{fig:angular}. After using smoothing splines, we differentiate $\Delta\theta(t)$ to obtain $\omega(t)$. In Fig.~\ref{fig:angular} we plot $\omega$ (normalized by the initial value $\omega_0$) versus time (normalized by the time of landing $t_f$). As expected, the angular velocity is perfectly constant for the rigid bottle (red markers). By contrast, $\omega$ is found to decrease dramatically for both the water bottle (blue) and the tennis bottle (yellow). These results reveal that a gentle landing can be achieved due to a significant reduction of the bottle's angular velocity $\omega$. The dashed lines in Fig.~\ref{fig:angular} correspond to the model developed in Sec.~\ref{sec:model}.

\subsection{Interpretation}

The secret behind a successful water bottle flip -- the reduced rotational velocity -- can be understood from conservation of angular momentum. The combined system of the bottle and the water is acted upon only by gravity, and therefore experiences no resultant torque around the center of mass. Consequently, the total angular momentum $L$ around the center of mass must be conserved: $L=I \omega$ is constant, where $I$ \textcolor{black}{is the moment of inertia relative to an axis passing through the center of mass.} The moment of inertia of a rigid body is constant over time, so that $\omega$ must remain constant -- in perfect agreement with our experiment (Fig.~\ref{fig:angular}, red markers). However, the mobility of the liquid gives rise to a redistribution of the mass inside the bottle during the flip, which implies that rotational inertia $I$ is no longer constant. This change of $I$ explains the reduction of the angular velocity observed in Fig.~\ref{fig:angular}: As the liquid mass spreads out, the total moment of inertia $I$ around the center of mass will increase, accompanied by a lowering of $\omega$ to maintain the same value of $L=I\omega$. The same argument applies for the tennis balls that ``spread out" during the flip, for which we indeed also observe a decrease in $\omega$.

\section{Model}\label{sec:model}

We now present a quantitative description of the experiments by modeling the redistribution of mass. \textcolor{black}{We first discuss the tennis bottle and subsequently propose a (highly simplified) one-dimensional model for the effect of sloshing inside the water bottle. In both cases we provide a quantitative comparison to experiments.} Finally, the model is used to address the question of what determines the optimal filling factor for a successful water bottle flip. 

% We first discuss the tennis bottle and subsequently propose a simplified one-dimensional model for the effect of sloshing inside the water bottle. In both cases we find good agreement with experiments. 

\subsection{The tennis bottle flip}

\subsubsection{Center of mass}

We start out by deriving the formula of the center of mass, which was already used for analyzing the experiment. The geometry of the tennis bottle is sketched in Fig.~\ref{fig:sketch}. 

The bottle is essentially a hollow cylinder of radius $R$, height $H$, and mass $m_b$. Assuming the cylinder is up-down symmetric, its center of mass is located at a position $H/2$. The tennis balls are modeled by hollow spheres of radius $R$ and mass $m_t$. The lower ball stays at the bottom of the cylinder while the top of the upper ball is located at a position $h$ that can change in the course of the experiment [cf. Fig.~\ref{fig:sketch}]. The center of mass of the two balls is thus located at $h/2$. 

The total center of mass of the combined system -- bottle and balls -- is obtained by a weighted average of the respective centers of mass. Hence, one verifies that the combined center of mass position $h_{\rm CM}$ is located at 

\begin{equation}\label{eq:hCMT}
h_{\rm CM} = \frac{m_b \frac{H}{2} + 2m_t \frac{h}{2}}{m_b + 2m_t} 
= \frac{H}{2}\left( \frac{m_b + 2 m_t \frac{h}{H}}{m_b + 2m_t }\right).
\end{equation}
Clearly, $h_{\rm CM}$ varies during the experiment, as it is a function of the position of $h$ of the second ball.

\subsubsection{Moment of inertia}

The next step is to determine the moment of inertia $I$ of the combined system. Given that the angular momentum is conserved only around the center of mass, we also need to determine $I$ with respect to the axis through the center of mass. 

Let us first consider the bottle. We assume that the bottle's mass is perfectly localized in a very thin wall at the outside of the cylinder (hence ignoring the mass in the top and bottom of the cylinder). With this, we can determine the moment of inertia in two steps. First, we consider the moment of inertia of the bottle with respect to the bottle's center of mass that is located at $H/2$, around the axis indicated in Fig.~\ref{fig:sketch}. This axis is perpendicular to the  cylinder's symmetry axis and the corresponding moment of inertia reads

\begin{equation}\label{eq:IBCM}
I'_{b} = \frac{m_b}{12}\left( 6R^2 + H^2 \right).
\end{equation}
However, the rotation takes place around the center of mass of the \emph{total} system $h_{\rm CM}$, defined by (\ref{eq:hCMT}). Hence, the axis of rotation in the experiment is parallel to the axis used for (\ref{eq:IBCM}), but shifted by a distance $\frac{H}{2} -h_{\rm CM}$. The relevant moment of inertia is then obtained by using the \emph{parallel axis theorem}. This gives

\begin{eqnarray}\label{eq:IB}
I_{b} &=& I_{b}' + m_b\left( \frac{H}{2} - h_{\rm CM}\right)^2 \nonumber \\
&=&\frac{m_b}{12}\left( 6R^2 + H^2 \right) + m_b\left( \frac{H}{2} - h_{\rm CM}\right)^2.
\end{eqnarray}

In a similar fashion, one obtains the moment of inertia of the two tennis balls. Approximating the balls as thin-walled hollow spheres, we obtain

\begin{eqnarray}
I_1 &=& \frac{2}{3}m_t R^2 + m_t\left( R - h_{\rm CM}\right)^2, \\
I_2 &=& \frac{2}{3}m_t R^2 + m_t\left( h-R - h_{\rm CM}\right)^2.
\end{eqnarray}
The first terms on the right hand side are the sphere's moment of inertia around its center of mass, while the second terms account for the parallel displacement to $h_{\rm CM}$ of the total system. 

Finally, the total moment of inertia during the tennis bottle flip reads 

\begin{equation}\label{eq:Itennis}
I(h) = I_b + I_1 + I_2.
\end{equation}
Each of these terms is a function of $h$, due to the dependence of $h_{\rm CM}$ on the position $h$ of the second ball. 

\subsubsection{Comparison to experiments}

To compare the model to experiments, we make use of the fact that the angular momentum around the center of mass, $L=I\omega$, must be conserved during the flip. According to this, we directly conclude that the dimensionless angular frequency $\omega(t)/\omega_0$ can be expressed as

\begin{equation}\label{eq:predictiontennis}
\frac{\omega(t)}{\omega_0} = \frac{I_0}{I(h)},
\end{equation}
with $I(h)$ given by (\ref{eq:Itennis}). Here we introduced the initial moment of inertia $I_0 = I(h_0)$, which corresponds to the situation prior to the flip when the two tennis balls are at the bottom of the container. Upon inspection of Fig.~\ref{fig:sketch} one finds $h_0=4R$. 

We now present two tests of our predictions. First, we insert the experimentally obtained $h(t)$ in (\ref{eq:Itennis}) and use (\ref{eq:predictiontennis}) to predict the angular velocity $\omega(t)$. The result is shown as the yellow dashed line in Fig.~\ref{fig:angular}. Clearly, it gives a very good description of the experimental data. 

\begin{figure}[ht]
\begin{center}
\includegraphics[width=.4\textwidth]{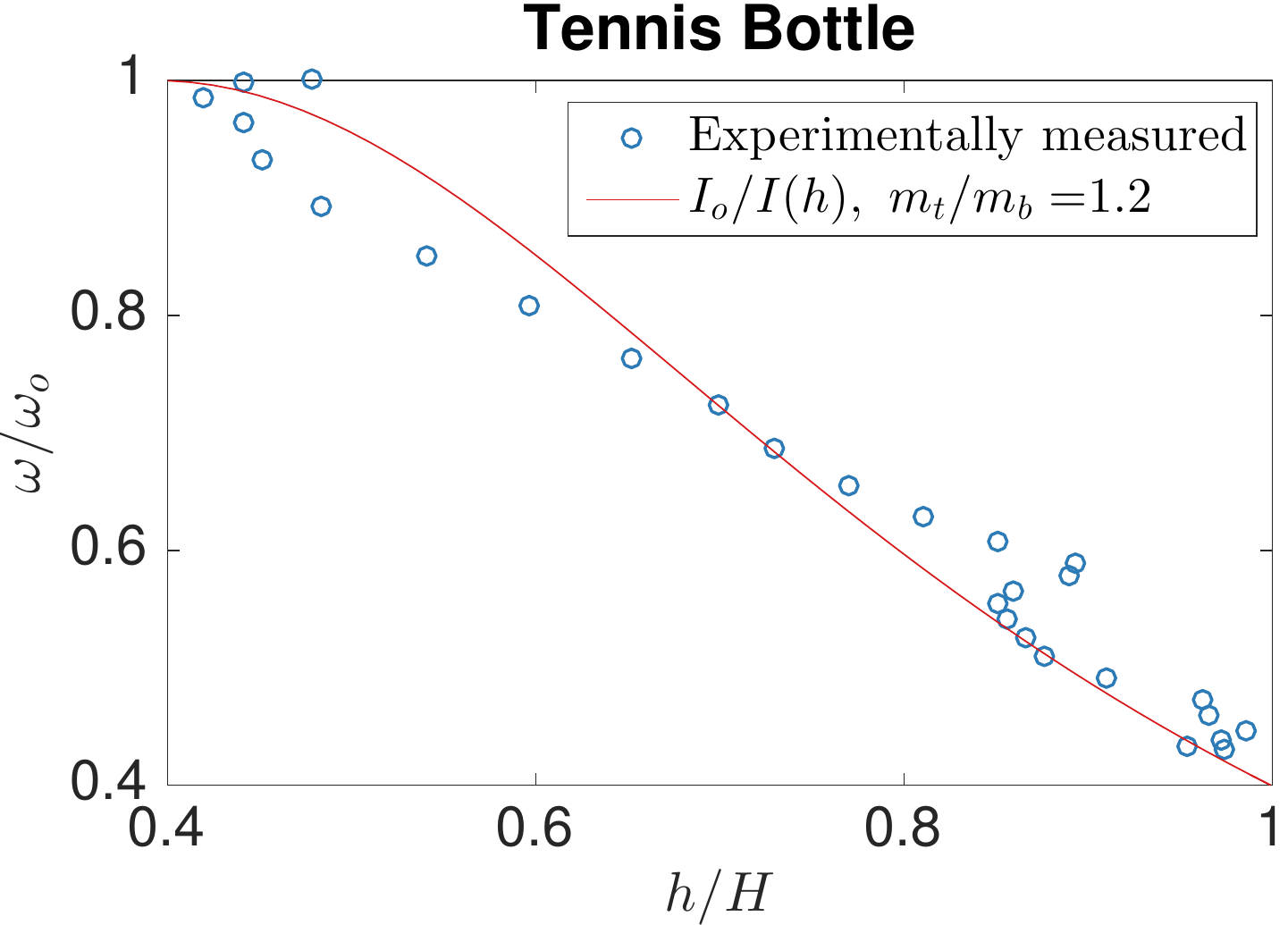}
\includegraphics[width=.4\textwidth]{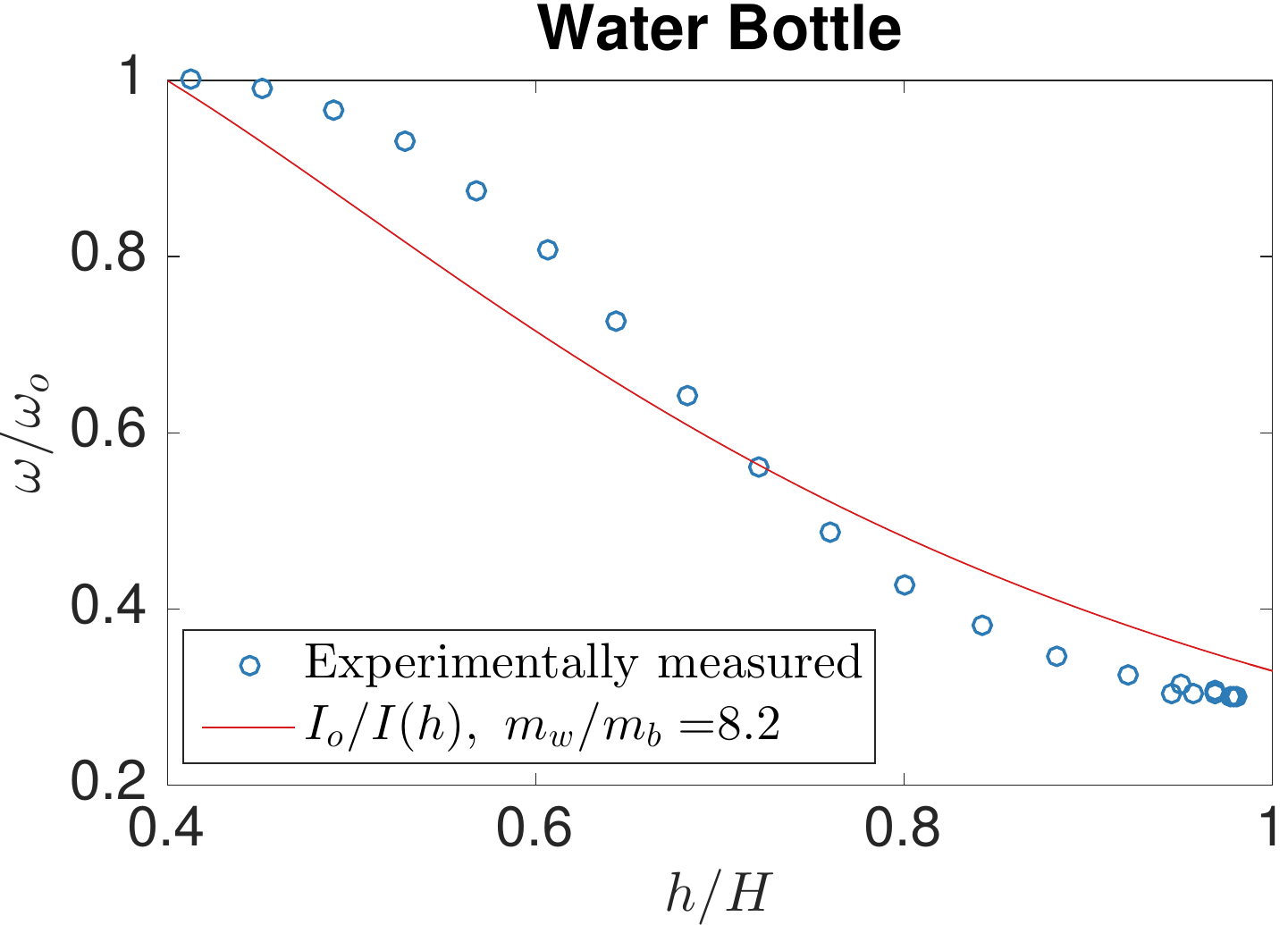}
\caption{Experimental results for the angular velocity $\omega$ for the tennis bottle (top panel) and the water bottle (bottom panel). Results of the models are shown in continuous red lines.} 
\label{fig:comparison}
\end{center}
\end{figure}

However, an even more direct verification of (\ref{eq:Itennis},\ref{eq:predictiontennis}) is obtained by plotting the experimental $\omega$ versus the experimental $h$. In this case the comparison between theory and experiment is without any input from experiment. The result is shown in Fig.~\ref{fig:comparison}, showing an excellent agreement without adjustable parameters. 

\subsection{The water bottle flip: \textcolor{black}{ a minimal }one-dimensional model}\label{subsec:modelwater}

\textcolor{black}{We now return to the case of the water bottle, for which the distribution of mass is obviously much more intricate. Our purpose here is not to provide a fully quantitative description of the sloshing fluid mass inside the bottle, which would require a detailed numerical treatment of the Navier-Stokes equations. Instead, we wish to propose a minimal model to that allows for a tractable approximate description of the water bottle flip. For this, we propose a simplified one-dimensional model.} We assume the mass of water $m_w$ is always distributed uniformly along the bottle, starting from the bottom and reaching up to a height $h$ (see Fig.~\ref{fig:sketch}). This height $h$ will vary with time as the bottle is spinning in the air. Again, we denote the minimum value of the height as $h_0$, which corresponds to the situation prior to the flip where all mass is collected at the bottom. The maximum possible value of $h$ is given by the height of the bottle $H$. \textcolor{black}{For simplicity, we further assume the mass to be distributed along the axis of the bottle}

% We now return to the case of the water bottle, for which the distribution of mass is obviously much more intricate. To describe the fluid mass in a tractable manner, without taking into account all the complexities of the sloshing fluid, we propose a simplified one-dimensional model. We assume the mass of water $m_w$ is always distributed uniformly along the bottle, starting from the bottom and reaching up to a height $h$ (see Fig.~\ref{fig:sketch}). This height $h$ will vary with time as the bottle is spinning in the air. Again, we denote the minimum value of the height as $h_0$, which corresponds to the situation prior to the flip where all mass is collected at the bottom. The maximum possible value of $h$ is given by the height of the bottle $H$. 

\subsubsection{Center of mass}

Once again, we first determine the center of mass of the combined system of the bottle (mass $m_b$) and the water (mass $m_w$). The center of mass can be found by taking the weighted average of the center of mass of the bottle, located at $H/2$, and of the distributed water, located at $h/2$. With this, the center of mass position can be found as 

\begin{equation}\label{eq:hcm}
    h_{\rm CM} = \frac{\frac{H}{2}m_b+\frac{h}{2}m_w}{m_b+m_w} = \frac{H}{2}\left(\frac{m_b + m_w \frac{h}{H}}{m_b+m_w} \right).
\end{equation}

This expression has been employed for obtaining the position of the CM in the water bottle experiments (see Figure \ref{fig:trajectories}).
% \textcolor{blue}{Was this used in the experimental determination of CM?}

%This expression already exhibits an interesting feature that is relevant for the water bottle flip. For an empty bottle the center of mass lies halfway the bottle, and the same holds true for a completely filled bottle. This can be verified by evaluating (\ref{eq:hcm}) respectively for $M=0$ (empty bottle) and $h/H=1$ (full bottle). In both cases one finds $h_{\rm CM}=H/2$. However, for intermediate filling factors the center of mass lies below the halfway point $H/2$. Hence, the position $h_{\rm CM}$ will vary during the experiment, as the water height evolves from the initial height $h_0$ to $H$. 

\subsubsection{Moment of inertia}

The next step is to determine the moment of inertia of the system $I$, measured with respect to the center of mass $h_{\rm CM}$. In analogy to the tennis bottle, we separately determine the moments of inertia of the bottle $I_b$ and of the water $I_w$, which leads to the total moment of inertia $I=I_b +I_w$. Using the parallel axis theorem, we find the bottle's moment of inertia to be

\begin{equation}
    I_b = I_0 + m_b\left(\frac{H}{2} - h_{\rm CM}\right)^2.
\end{equation}

Here $I_0$ \textcolor{black}{is the moment of inertia of the bottle relative to an axis passing through its own center of mass }(located approximately at $H/2$), while the second term accounts for the shift to the system's center of mass at $h_{\rm CM}$. Since we consider a simplified one-dimensional description, we will from now on use $I_0=\frac{1}{12}m_b H^2$. 
\textcolor{black}{This is the expression valid for thin objects where all mass is located along the axis, and is also recovered from (\ref{eq:IBCM}) with $R = 0$. Equation (\ref{eq:IBCM}) in fact allows for an estimation of the correction induced from the fact the water mass is not on the axis, the relative error being $6R^2/H^2$, which yields an error about 12\% for the water bottle used in our study ($R/H\approx1/7$)}. In similar fashion, we can express the moment of inertia of the one-dimensional water column as

% This is the expression valid for thin objects, and is also recovered from (\ref{eq:IBCM}) with $R=0$. 

\begin{equation}
    I_w = \frac{1}{12}m_w h^2 + M\left(\frac{h}{2} - h_{\rm CM}\right)^2.
\end{equation}
The total moment of inertia then reads

\begin{eqnarray}\label{eq:Itot}
I = I_b + I_w = \frac{1}{12}\left( m_bH^2 + m_wh^2 \right) \nonumber \\
 + m_b\left(\frac{H}{2} - h_{\rm CM}\right)^2 + m_w\left(\frac{h}{2} - h_{\rm CM}\right)^2,
\end{eqnarray}
where it is understood that $h_{\rm CM}$ is given by (\ref{eq:hcm}). 

\subsubsection{Comparison to experiments}

We now make the same comparison to the experiments as we did for the tennis bottle. This is again based on 

\begin{equation}
\frac{\omega(t)}{\omega_0} = \frac{I_0}{I(h)},
\end{equation}
but now with $I(h)$ based on (\ref{eq:Itot}). 

The first comparison is shown as the blue dashed line in Fig.~\ref{fig:angular}, where we used $h(t)$ measured in the experiment.  The same data are shown Fig.~\ref{fig:comparison}, now plotting $\omega$ versus $h$. 
\textcolor{black}{Without any adjustable parameters, the model gives a reasonable account of the reduction of $\omega$ during water bottle flips. In particular given the oversimplification of the sloshing in this one-dimensional description. Some of the features, such as the appearance of an inflection point halfway the flip, are not captured, which could be due to the fact that the mass does not remain distributed along the central axis of the bottle.}

% The model gives a very good account of the reduction of $\omega$ during water bottle flips, in particular given the oversimplification of the sloshing in this one-dimensional description.

\section{Can we predict the optimal filling fraction?}\label{sec:optimal}

Encouraged by these observations, we now turn to the question of what is the optimal filling fraction, $f=h_0/H$, to accomplish a water bottle flip. It is obvious that an optimum should exist. Namely, both an empty bottle ($f=0$) and a filled bottle ($f=1$) can not accommodate any mass redistribution, and hence will not exhibit any slowing down of $\omega$. According to the model, What would be the optimal $f$?

\subsection{Reducing the angular velocity}

\begin{figure}[h]
\begin{center}
\includegraphics[width=.5\textwidth]{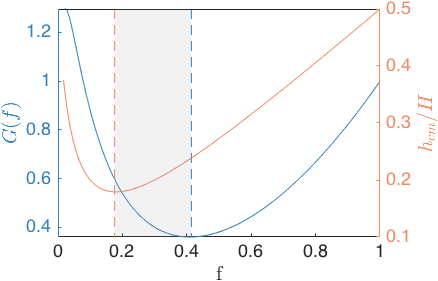}
\caption{Two criteria for the \emph{optimal flip}  are shown. On the left y-axis, the minimum in $G(f)$ (i.e. maximum moment of inertia increase $I/I_o$) gives us the first criterion for an optimal bottle filling fraction $f$. On the right y-axis, the lowest achievable center of mass position gives us another criterion for an \emph{optimal flip}. The plot shows the case of bottle water/bottle mass ratio $M=m_{\rm water max}/m_{\rm bottle}$= 20. }
\label{fig:optimal}
\end{center}
\end{figure}

Since for a given $\omega_0$ one wishes to reduce $\omega$ as much as possible, we will look for the minimum of the ratio $I_0/I(h)$. For each filling fraction, the maximum moment of inertia $I_{\rm max}$ is attained when the water is maximally distributed, i.e. for $h=H$. Hence, we need to find the value of $f$ for which $I_0/I_{\rm max}$ attains a minimum. Though the expression (\ref{eq:Itot}) appears rather cumbersome, it is possible to find an analytical form for the function $G(f)\equiv I_0/I_{\rm max}$. For this we first define the mass ratio 

\begin{equation}
M =\frac{m_{w,{\rm max}}}{m_b},
\end{equation}
where $m_{w,{\rm max}}$ is the water mass for a filled bottle. With this, we can express $m_w = f m_{w,{\rm max}} = f M m_b$ and insert this in (\ref{eq:Itot}) and (\ref{eq:hcm}). With the help of Maple \cite{Maple} or Mathematica \cite{Mathematica}, the remaining expression can be brought to the form

\begin{equation}
G(f) = \frac{I_0}{I_{\rm max}} = \frac{M^2 f^4 + 4Mf^3 - 6Mf^2 + 4Mf + 1}{\left( 1 + Mf \right)^2}.
\end{equation}
This relation is plotted in Fig.~\ref{fig:optimal} (blue curve) for M=20. Typically, water bottles that can contain 0.5 liters of water have a mass of approximately 25 grams. This implies $M=500/25=20$, for which $G(f)$ exhibits a minimum at $f\approx 0.41$. The corresponding reduction $\omega_{\rm min}/\omega_0 \approx 0.36$, which we remark to be in close agreement with the reduction achieved experimentally in Fig.~\ref{fig:angular} (in our experiments $f=0.39$).

\subsection{Lowering the center of mass}

One might argue that the optimization involves more than just the reduction of $\omega$. After all, the dynamics of the landing is also of key importance. Clearly, the stability of the landing would benefit from having the center of mass as low as possible. Another relevant minimization would therefore be $h_{\rm CM}$ evaluated for $h=h_0$. Again introducing the mass ratio $M$, the expression (\ref{eq:hcm}) can be written as

\begin{equation}
\frac{h_{\rm CM}}{H} = \frac{1}{2}\left( \frac{1+Mf^2}{1+Mf} \right).
\end{equation}
This result is shown in Fig.~\ref{fig:optimal} (red curve), again for $M=20$. Now, the minimization with respect to $f$ can be performed analytically and yields

\begin{equation}
f = \frac{\sqrt{1+M}-1}{M}.
\end{equation}
For $M=20$ this gives $f=0.18$. 

Using the these two criteria of having a low angular velocity and a low center of mass, our crude model provides a prediction for the optimal range. This is shown as the gray zone in Fig.~\ref{fig:optimal}. The figure shows that good filling fractions lie in the range of approximately 20\% to 40\%. This is consistent with the reports found on the Internet, which typically quote 1/4 to 1/3.

\section{Discussion}

To summarize, we have presented the physics of the water bottle flip as a contemporary illustration of the principles of rotational mechanics. It allows for a variety of experimental and theoretical explorations that are suitable to undergraduate physics courses. In fact, the research presented here was initiated, and to a large extent executed, by the five undergraduate students who appear as the first authors of this paper.\cite{Footnote1} Possible extensions of the presented work are to investigate the role of horizontal momentum for a successful landing, or to analyze the landing itself.

Apart from its intrinsic interest, the principle of redistribution of mass finds applications in a variety of contexts. For example, Olympic divers extend their arms and legs as much as possible to reduce their rotational speed and dive into the water in a straight position. Similar strategies are used in granular dampers, in which solid particles inside a shaky object are used to damp undesired oscillations and stabilize the object \cite{dampers}. These examples give a broader perspective on the physics behind the water bottle flip. 

%\begin{acknowledgments}
%We wish to thank...
%\end{acknowledgments}

% \bibliographystyle{unsrt}
% \bibliography{bibbottle.bib}

\begin{thebibliography}{99}
\bibitem{Footnote1}
The research project was part of a course on classical mechanics, in the first year of the BSc program on Applied Physics at the University of Twente.
\end{thebibliography}

\end{document}